\documentclass[onecolumn,citeautoscript,superscriptaddress,notitlepage]{revtex4-1}     
\usepackage{amsmath,amssymb,mathrsfs,bm,setspace,xspace,soul,empheq}  
\usepackage{graphicx}
\usepackage{float}  
\usepackage{comment}           
\usepackage[tight]{subfigure}                   
\usepackage{braket}                 
\usepackage{color}
\usepackage[colorinlistoftodos]{todonotes}             
\usepackage{tikz} 
\usetikzlibrary{shapes,decorations,calc}       
\usepackage{dcolumn}
\usepackage{multirow}             
\usepackage{geometry}              
\usepackage{tabularx}    
\usepackage[colorlinks=true]{hyperref}      
\hypersetup{ 
    bookmarks=true,         
    unicode=false,          
    pdftoolbar=true,        
    pdfmenubar=true,        
    pdffitwindow=false,     
    pdfstartview={FitH},    
    pdftitle={My title},    
    pdfauthor={Author},     
    pdfsubject={Subject},   
    pdfcreator={Creator},   
    pdfproducer={Producer}, 
    pdfkeywords={keyword1} {key2} {key3}, 
    pdfnewwindow=true,      
    colorlinks=true,       
    linkcolor=magenta, 
    citecolor=blue,        
    filecolor=magenta,      
    urlcolor=cyan           
} 

\renewcommand{\ket}[1]{\left| #1\right\rangle}

\newcommand{\beq}{\begin{equation}}
\newcommand{\eeq}{\end{equation}}
\newcommand{\ba}{\begin{array}{ccc}}
\newcommand{\ea}{\end{array}}

\def\bea{\begin{eqnarray}}
\def\eea{\end{eqnarray}}
\newcommand{\bml}{\begin{multline}}

\newcommand{\eeqm}{\end{multline}}
\newcommand{\bsp}{\begin{split}}
\newcommand{\esp}{\end{split}}

\renewcommand{\b}[1]{{\bf #1}}

\newcommand{\approptoinn}[2]{\mathrel{\vcenter{
  \offinterlineskip\halign{\hfil$##$\cr
    #1\propto\cr\noalign{\kern2pt}#1\sim\cr\noalign{\kern-2pt}}}}}

\newcommand{\cf}{\mathfrak b}
\newcommand{\cffl}{\cf_{\rm FL}}

\begin{document}  

\title{Cornering the universal shape of fluctuations}   

\author{Benoit Estienne}
\affiliation{Sorbonne Universit\'{e}, CNRS, Laboratoire de   Physique Th\'{e}orique et Hautes \'{E}nergies, LPTHE, F-75005 Paris, France.}
\author{Jean-Marie St\'ephan}
\affiliation{Univ Lyon, CNRS, Universit\'e Claude Bernard Lyon 1, UMR5208, Institut Camille Jordan, F-69622 Villeurbanne, France}
\author{William Witczak-Krempa}
\affiliation{D\'epartement de Physique, Universit\'e de Montr\'eal, Montr\'eal, Qu\'ebec, H3C 3J7, Canada}
\affiliation{Centre de Recherches Math\'ematiques, Universit\'e de Montr\'eal; P.O. Box 6128, Centre-ville Station; Montr\'eal (Qu\'ebec), H3C 3J7, Canada}
\affiliation{Regroupement Qu\'eb\'ecois sur les Mat\'eriaux de Pointe (RQMP)}

\begin{abstract}  
\begin{center}
\textbf{Abstract}
\end{center}
Understanding the fluctuations of observables is one of the main goals in science, be it theoretical or experimental, quantum or classical. 
We investigate such fluctuations when only a subregion of the full system can be observed, focusing on geometries with sharp corners. We report that the dependence on the opening angle is \emph{super-universal}: up to a numerical prefactor, this function does not depend on anything, provided the system under study is uniform, isotropic, and correlations do not decay too slowly. The prefactor contains important physical information: we show in particular that it gives access to the long-wavelength limit of the structure factor. We illustrate our findings with several examples, including fractional quantum Hall states, scale invariant quantum critical theories, and metals.
Finally, we discuss connections with quantum entanglement, extensions to three dimensions, as well as experiments to probe the geometry of fluctuations.  
\end{abstract}
    
\date{\today}     
\maketitle            

\section{Universal shape of fluctuations}   
In quantum mechanics, measurements on identically prepared systems of an observable $\mathcal O$
will generally yield different outcomes. This is a consequence of the fact that the state of the system is in a quantum superposition of states having well-defined values of $\mathcal O$. 
The spread of the outcomes, ignoring experimental errors, can be quantified by the variance, or uncertainty squared in the quantum language,
$(\Delta\mathcal O)^2=\langle (\mathcal O-\langle \mathcal O\rangle)^2\rangle$.
Heuristically, we say that $\Delta\mathcal O$ measures the fluctuations of $\mathcal O$. Similar fluctuations also occur in classical many-body systems,
where the statistical description leads to fluctuations of observables.
In numerous experiments, like scanning tunneling microscopy, one only measures a small subregion of a sample.  
In that case, a natural question arises: What are the fluctuations of a given observable in a subregion $A$?
This refinement introduces additional information: the shape of the subregion. It thus seems that one is left with a huge amount of possibilities corresponding to different quantum or classical states, observables, and shapes, and thus little hope to find unifying principles.
In this work we show that there exists a large, and experimentally relevant, set of states and observables that share the same universal shape dependence for their fluctuations.   
 
Let us consider a local scalar observable, written in the continuum as $\rho(\b r)$. It could be the number of bacteria per unit area, the charge density, the energy density,
the local magnetization, etc. The fluctuations of $\rho$ within a subregion $A$ are described by $\Delta\mathcal O_A$, where $\mathcal O_A =\int_A d\b r\,  \rho({\bf r})$
is the integrated density in the subregion:
\begin{align}
  (\Delta \mathcal O_A)^2 = \langle \mathcal O_A^2\rangle - \langle \mathcal O_A\rangle^2
  = \int_A d\bm r \int_A d\bm r' \langle \rho(\bm r) \rho(\bm r')\rangle_c
\end{align}
with the connected correlation function
$ \langle \rho(\bm r)   \rho(\bm r')\rangle_c = \langle \rho(\bm r) \rho(\bm r')\rangle - \langle  \rho(\bm r) \rangle\langle \rho(\bm r')\rangle $.
The expectation value is taken either with respect to a classical distribution, or a quantum density matrix. We will now focus on uniform and isotropic systems, for which the above correlation function only depends on the distance separating the two positions $ \langle \rho(\bm r)   \rho(\bm r')\rangle_c = f(| \bm r-\bm r'|)$, yielding  
\begin{align}
  (\Delta \mathcal O_A)^2 = \int_A d\bm r \int_A d\bm r'\, f(| \bm r-\bm r'|)\, .
\end{align}
The function $f$ can be very different depending on the system and choice of observable.
From general principles, fluctuations of most physical systems behave for large regions $A$ as 
\begin{align}
  (\Delta \mathcal O_A)^2  =\alpha |A| +\beta |\partial A| -  b_A + \cdots \,.
\end{align}
The first term is a standard volume law, scaling with the size of $A$, while the second term is an area law scaling with the size of its 
boundary $\partial A$.  
The prefactors $\alpha$ and $\beta$ do not depend on the shape of region $A$, and they can be explicitly computed in terms of the correlation function $f$ (see Supplementary Material).
The subleading term $b_A$ is more interesting : it carries the non-trivial shape dependence of the fluctuations and probes the  large-scale properties of the system.
In particular, if $A$ has sharp corners, each one contributes to $b_A$. These corner contributions are encoded in a function $b(\theta)$, $\theta$ being
the corner opening angle. The case of a simple planar corner in two dimensions is illustrated in Fig.~1.

For the sake of concreteness, we shall now turn to the important case of two spatial dimensions. We report that the angle-dependence $b(\theta)$ of the fluctuations is in fact completely independent of the observable and of the system considered, up to a numerical prefactor. Namely 
\begin{align} \label{eq:master}
  b(\theta)
  = -  \big( 1+(\pi -\theta)\cot\theta \big) \int_0^\infty \frac{r^3}{2} f(r)\, dr
\end{align} 
as long as the system is translation invariant and isotropic, and the correlation function $f$ decays
  sufficiently fast at large $r$, as discussed below. 
This is the main result of this paper. We emphasize that the aforementioned assumptions have considerable generality: they hold for a wide class of classical and  quantum systems, at zero or finite temperature. A typical example would be that of a liquid, where in addition to translational invariance and isotropy, $f$ decays exponentially fast.
A generalization to three dimensions shall be discussed at the end.  

Strikingly, the simple angle dependence factorizes and is independent from the correlation function $f$. 
Before we provide the derivation of this result and present several non-trivial tests, it is worthwhile to pause and examine the angular function in Eq.~(\ref{eq:master}),
$u(\theta)= 1+(\pi-\theta)\cot\theta$,
which we call the \emph{corner fluctuation} function. It is plotted in Fig.~\ref{fig:fqhnumerics} (left).
Due to the appearance of the cotangent, $\cot\theta=\tfrac{\cos\theta}{\sin\theta}$, it diverges as $1/\theta$ when the angle approaches zero.
The increase at small angles is natural given that the region is becoming thinner, which leads to stronger long-range fluctuations. 
In the opposite limit of $\theta\approx\pi$, it vanishes quadratically as $(\theta-\pi)^2$. 

The prefactor of the corner fluctuation function, that is the radial integral in Eq.~(4), is also meaningful and holds interesting physical information. 
This coefficient can  be measured experimentally, as it is directly related to the long-wavelength limit of the static structure factor,
which can be accessed via elastic scattering experiments, for example. We shall treat various examples below. The reader may remark that the integral is free from large-scale divergence provided $f$ decays faster than $1/r^4$. When the decay is precisely $1/r^4$, one obtains a logarithmic divergence with the size of region $A$, as we shall explain when we treat scale-invariant quantum critical systems. However, in certain situations the decay is even slower, and we will obtain a new scaling.  

The corner fluctuation function has previously appeared in several contexts, for example in renormalization studies of Wilson
loops in gauge theories~\cite{Brandtetal1981,Korchemsky1987}, in the study of entanglement entropy~\cite{Casini2005,Casini2008,Swingle2010}, and       
in bipartite fluctuations of non-interacting Dirac systems \cite{Herviouetal2019} and the integer quantum Hall effect \cite{estienne2019entanglement}.
Our findings illuminate its physical origin in large class of classical and quantum systems, and explain why it has appeared in these seemingly unrelated contexts.

The paper is organized as follows. In Sec.~\ref{sec:derivation} we present a general derivation of our main result. The rest of the paper focuses on examples, which illustrate its ubiquity. Sec.~\ref{sec:fqh} deals with fractional quantum Hall systems, where even the prefactor of the corner term is universal and proportional to the Hall conductivity. This example can also be interpreted as a classical (liquid) particle system with 2d Coulomb repulsion via the plasma analogy.  We turn in Sec.~\ref{sec:cft} to quantum critical
scale-invariant theories, for which the corner function diverges logarithmically: the prefactor is also universal in that case, but has a different physical origin. It is proportional to the longitudinal conductivity. We investigate the case of metals in Sec.~\ref{sec:metals}, which breaks our assumptions and shows different behavior. Finally,
we discuss numerous ramifications and extensions of our analysis in Sec.~\ref{sec:disc}, including to higher dimensions.

\section{Obtaining the universal corner function}\label{sec:derivation} 

\begin{figure}[htbp]
 \centering\begin{tikzpicture}[scale=0.7]
 \fill[red,opacity=0.5] (0,0) -- (4.5cm,0mm) arc (0:65:4.5cm) -- (0,0);
 \begin{scope}[opacity=0.6]
 \filldraw[green!40!black] (-0.72925,-4.42225) circle (0.1cm);
\filldraw[green!40!black] (-3.71275,0.467) circle (0.1cm);
\filldraw[green!40!black] (1.099,3.43475) circle (0.1cm);
\filldraw[green!40!black] (-2.6165,-1.5875) circle (0.1cm);
\filldraw[green!40!black] (2.50025,0.05225) circle (0.1cm);
\filldraw[green!40!black] (1.86575,-0.1765) circle (0.1cm);
\filldraw[green!40!black] (-3.91525,1.15925) circle (0.1cm);
\filldraw[green!40!black] (3.21725,0.58475) circle (0.1cm);
\filldraw[green!40!black] (-3.17225,-2.05825) circle (0.1cm);
\filldraw[green!40!black] (0.40275,-2.14225) circle (0.1cm);
\filldraw[green!40!black] (1.89075,-1.28775) circle (0.1cm);
\filldraw[green!40!black] (1.45325,-0.05175) circle (0.1cm);
\filldraw[green!40!black] (-3.93925,-0.694) circle (0.1cm);
\filldraw[green!40!black] (1.39775,2.96675) circle (0.1cm);
\filldraw[green!40!black] (3.26475,-1.0815) circle (0.1cm);
\filldraw[green!40!black] (-2.0625,0.253) circle (0.1cm);
\filldraw[green!40!black] (-2.84575,1.756) circle (0.1cm);
\filldraw[green!40!black] (3.7945,-0.682) circle (0.1cm);
\filldraw[green!40!black] (-1.70125,-3.50875) circle (0.1cm);
\filldraw[green!40!black] (0.42825,3.83825) circle (0.1cm);
\filldraw[green!40!black] (2.679,-0.98025) circle (0.1cm);
\filldraw[green!40!black] (0.18475,-3.0595) circle (0.1cm);
\filldraw[green!40!black] (0.11875,0.7355) circle (0.1cm);
\filldraw[green!40!black] (-1.9525,2.95) circle (0.1cm);
\filldraw[green!40!black] (0.608,-1.1765) circle (0.1cm);
\filldraw[green!40!black] (2.73225,3.13975) circle (0.1cm);
\filldraw[green!40!black] (-0.861,1.06075) circle (0.1cm);
\filldraw[green!40!black] (-3.3445,1.30625) circle (0.1cm);
\filldraw[green!40!black] (1.10875,3.8505) circle (0.1cm);
\filldraw[green!40!black] (0.5175,2.642) circle (0.1cm);
\filldraw[green!40!black] (-1.56225,-1.69525) circle (0.1cm);
\filldraw[green!40!black] (0.65025,0.8) circle (0.1cm);
\filldraw[green!40!black] (3.6175,1.27225) circle (0.1cm);
\filldraw[green!40!black] (0.7825,-2.358) circle (0.1cm);
\filldraw[green!40!black] (1.2655,-2.39275) circle (0.1cm);
\filldraw[green!40!black] (-2.785,3.2255) circle (0.1cm);
\filldraw[green!40!black] (-1.2985,-3.2425) circle (0.1cm);
\filldraw[green!40!black] (-1.98275,-2.96575) circle (0.1cm);
\filldraw[green!40!black] (-0.2315,1.49425) circle (0.1cm);
\filldraw[green!40!black] (-0.9185,-2.1805) circle (0.1cm);
\filldraw[green!40!black] (-1.93825,-1.53925) circle (0.1cm);
\filldraw[green!40!black] (-0.1895,-0.22975) circle (0.1cm);
\filldraw[green!40!black] (-1.03,1.43275) circle (0.1cm);
\filldraw[green!40!black] (-0.236,2.13425) circle (0.1cm);
\filldraw[green!40!black] (1.7365,-2.234) circle (0.1cm);
\filldraw[green!40!black] (-1.306,2.07125) circle (0.1cm);
\filldraw[green!40!black] (0.2865,-4.02025) circle (0.1cm);
\filldraw[green!40!black] (-1.30275,0.0315) circle (0.1cm);
\filldraw[green!40!black] (-0.305,-1.174) circle (0.1cm);
\filldraw[green!40!black] (-2.98125,0.9765) circle (0.1cm);
\filldraw[green!40!black] (-2.91775,-0.52) circle (0.1cm);
\filldraw[green!40!black] (-1.43875,0.831) circle (0.1cm);
\filldraw[green!40!black] (1.59425,-2.849) circle (0.1cm);
\filldraw[green!40!black] (-1.94675,-0.434) circle (0.1cm);
\filldraw[green!40!black] (-1.3685,3.10225) circle (0.1cm);
\filldraw[green!40!black] (-3.31975,-0.16575) circle (0.1cm);
\filldraw[green!40!black] (-2.8225,0.16525) circle (0.1cm);
\filldraw[green!40!black] (0.8985,-0.4265) circle (0.1cm);
\filldraw[green!40!black] (-1.25,2.522) circle (0.1cm);
\filldraw[green!40!black] (3.34325,1.991) circle (0.1cm);
\filldraw[green!40!black] (2.73775,-1.52075) circle (0.1cm);
\filldraw[green!40!black] (2.96775,1.51525) circle (0.1cm);
\filldraw[green!40!black] (-0.48,4.02325) circle (0.1cm);
\filldraw[green!40!black] (-0.69825,3.245) circle (0.1cm);
\filldraw[green!40!black] (-2.95925,-2.74325) circle (0.1cm);
\filldraw[green!40!black] (-4.105,-0.18075) circle (0.1cm);
\filldraw[green!40!black] (2.0575,0.656) circle (0.1cm);
\filldraw[green!40!black] (0.31225,2.1905) circle (0.1cm);
\filldraw[green!40!black] (-2.34025,-3.269) circle (0.1cm);
\filldraw[green!40!black] (1.871,1.64925) circle (0.1cm);
\filldraw[green!40!black] (-2.35,0.99425) circle (0.1cm);
\filldraw[green!40!black] (1.06825,-3.648) circle (0.1cm);
\filldraw[green!40!black] (-1.50675,3.77175) circle (0.1cm);
\filldraw[green!40!black] (2.1545,2.91675) circle (0.1cm);
\filldraw[green!40!black] (-2.181,2.13625) circle (0.1cm);
\filldraw[green!40!black] (-0.98475,2.775) circle (0.1cm);
\filldraw[green!40!black] (-0.3465,3.561) circle (0.1cm);
\filldraw[green!40!black] (1.55375,-3.70675) circle (0.1cm);
\filldraw[green!40!black] (2.2775,2.07825) circle (0.1cm);
\filldraw[green!40!black] (-1.2885,-1.15625) circle (0.1cm);
\filldraw[green!40!black] (-1.989,1.473) circle (0.1cm);
\filldraw[green!40!black] (-2.404,-1.11675) circle (0.1cm);
\filldraw[green!40!black] (-1.00525,-1.8155) circle (0.1cm);
\filldraw[green!40!black] (0.8115,-2.957) circle (0.1cm);
\filldraw[green!40!black] (1.8885,-0.7925) circle (0.1cm);
\filldraw[green!40!black] (2.261,-2.80275) circle (0.1cm);
\filldraw[green!40!black] (2.9115,2.46375) circle (0.1cm);
\filldraw[green!40!black] (3.07225,-1.99825) circle (0.1cm);
\filldraw[green!40!black] (0.3595,-3.319) circle (0.1cm);
\filldraw[green!40!black] (3.68325,0.07175) circle (0.1cm);
\filldraw[green!40!black] (-0.53475,-3.84625) circle (0.1cm);
\filldraw[green!40!black] (-3.2865,2.04) circle (0.1cm);
\filldraw[green!40!black] (-3.869,-1.289) circle (0.1cm);
\filldraw[green!40!black] (1.23925,1.4525) circle (0.1cm);
\filldraw[green!40!black] (-0.201,0.16425) circle (0.1cm);
\filldraw[green!40!black] (-0.81425,-1.052) circle (0.1cm);
\filldraw[green!40!black] (1.84575,3.39125) circle (0.1cm);
\filldraw[green!40!black] (-1.46025,-2.18225) circle (0.1cm);
\filldraw[green!40!black] (2.65275,-0.3625) circle (0.1cm);
\filldraw[green!40!black] (-0.16525,2.702) circle (0.1cm);
\filldraw[green!40!black] (3.62375,-1.5635) circle (0.1cm);
\filldraw[green!40!black] (3.8495,0.286) circle (0.1cm);
\filldraw[green!40!black] (-3.26425,-0.7715) circle (0.1cm);
\filldraw[green!40!black] (-0.73725,-0.584) circle (0.1cm);
\filldraw[green!40!black] (-1.769,0.56475) circle (0.1cm);
\filldraw[green!40!black] (-0.9905,-3.51375) circle (0.1cm);
\filldraw[green!40!black] (1.0695,-0.6905) circle (0.1cm);
\filldraw[green!40!black] (-3.2135,-1.428) circle (0.1cm);
\filldraw[green!40!black] (2.16875,-1.58575) circle (0.1cm);
\filldraw[green!40!black] (0.06725,1.3235) circle (0.1cm);
\filldraw[green!40!black] (1.4805,2.04925) circle (0.1cm);
\filldraw[green!40!black] (2.91275,1.02925) circle (0.1cm);
\filldraw[green!40!black] (-2.82925,2.54675) circle (0.1cm);
\filldraw[green!40!black] (3.16625,0.0375) circle (0.1cm);
\filldraw[green!40!black] (-0.48475,-2.79875) circle (0.1cm);
\filldraw[green!40!black] (-2.30825,-1.96925) circle (0.1cm);
\filldraw[green!40!black] (0.16125,-0.7295) circle (0.1cm);
\filldraw[green!40!black] (1.134,-1.6705) circle (0.1cm);
\filldraw[green!40!black] (0.2805,3.161) circle (0.1cm);
\filldraw[green!40!black] (0.7025,0.0835) circle (0.1cm);
\filldraw[green!40!black] (0.961,2.47725) circle (0.1cm);
\filldraw[green!40!black] (1.68275,1.11725) circle (0.1cm);
\filldraw[green!40!black] (-0.06075,-2.1835) circle (0.1cm);
\filldraw[green!40!black] (1.337,0.56975) circle (0.1cm);
\filldraw[green!40!black] (-0.77325,0.52075) circle (0.1cm);
\filldraw[green!40!black] (2.55375,-3.008) circle (0.1cm);
\filldraw[green!40!black] (2.718,-2.63325) circle (0.1cm);
\filldraw[green!40!black] (2.12475,1.6355) circle (0.1cm);
 \end{scope}
  \draw[very thick] (0,0) -- (4,0);
 \draw[very thick,rotate=65] (0,0) -- (4,0);
 \draw[->,ultra thick,blue!60!black] (1.6,0) arc (0:65:1.6); 
 \draw[blue!60!black] (1.75,0.7) node {\LARGE{$\theta$}};
\end{tikzpicture}
\hspace{2cm}
\begin{tikzpicture}[scale=0.7]
 \fill[red,opacity=0.5] (0,0) -- (4.5cm,0mm) arc (0:65:4.5cm) -- (0,0);
  \draw[very thick] (-4,0) -- (4,0);
 \draw[very thick,rotate=65] (-4,0) -- (4,0);
 \draw (2.4,1.1) node {\LARGE{$A$}};
 \draw (-0.9,1.8) node {\LARGE{$B$}};
 \draw (-1.9,-0.9) node {\LARGE{$C$}};
 \draw (1.5,-1.7) node {\LARGE{$D$}};
 \draw[->,ultra thick,blue!60!black] (1.6,0) arc (0:65:1.6); 
 \draw[blue!60!black] (1.75,1) node {\LARGE{$\theta$}};
\end{tikzpicture}
\caption{{\bf Corner geometry.} {\bf Left:} Subregion $A$ is highlighted in red. The electron distribution is a typical Monte Carlo sample obtained
    from the topological fractional quantum Hall groundstate at filling $\nu=1/3$.
    {\bf Right:} The regions that are used in the substraction procedure for cancelling out the boundary law, and isolating the corner contribution. 
    } 
 \label{fig:corner} 
\end{figure}
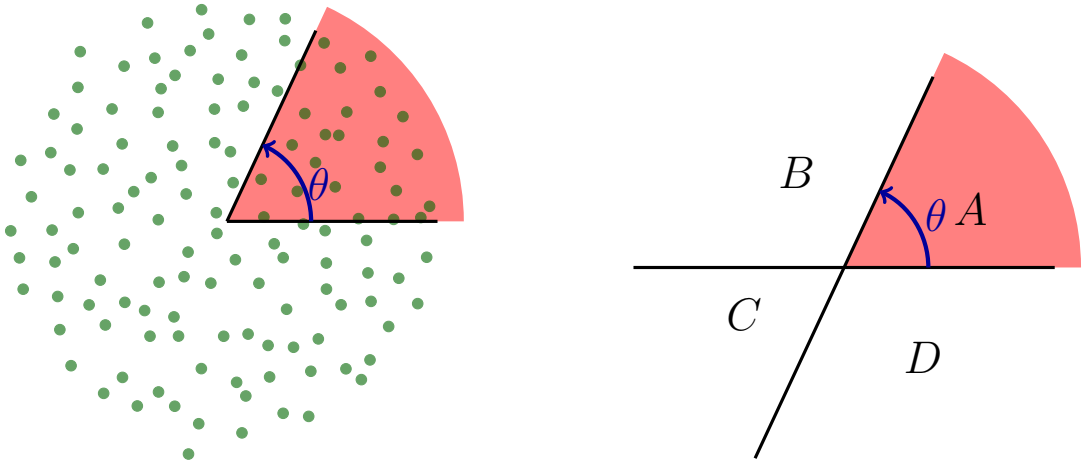

In order to evaluate the corner contribution $b(\theta)$, we consider for the region $A$ a single corner of opening angle $\theta$, as illustrated in Fig.~\ref{fig:corner} (left). 
First, we must isolate the subleading corner contribution from the volume and boundary terms. The volume law is easy to take care of. Its origin can be traced back to the fluctuations of the total integrated density  
$\mathcal{O} = \int_{V} d{\bf r} \rho({\bf r})$. Indeed, unless $\mathcal O$ is conserved, its variance is extensive with the total system size $V$, with a density of fluctuations (see e.g.\ \cite{Songetal2012}) $\alpha =  (\Delta \mathcal O)^2/V  = \int_{} d{\bf r} \braket{ \rho({\bf r}) \rho({\bf 0})}_c $.
Straightforward manipulations lead to 
\begin{align} \label{eq:comp}
  (\Delta \mathcal O_A)^2 =   \alpha |A| -  \int_A d\bm r \int_{A^c} d\bm r'\, f(| \bm r-\bm r'|)
\end{align}
where $A^c$ denotes the complement of region $A$. This takes care of the volume term, the second term in the \emph{r.h.s.} being governed by an area law provided $f$ decays sufficiently fast to zero.  
In particular if $\mathcal{O}$ does not fluctuate,  the volume contribution vanishes \footnote{Care must be taken at finite temperature in the canonical ensemble. In this case $\alpha$ will vanish but the function $f$ will not decay to zero, resulting in a separate volume term. This can be easily seen at infinite temperature, where correlations do not quite vanish due to the constraint on particle number in the whole system. In contrast, $\alpha$ is the only possible volume law contribution in the grand-canonical ensemble.}. This is for example the case for ground states of local Hamiltonians that respect the symmetry corresponding to $\mathcal{O}$. In the appropriate temperature regime, the fluctuations of $\mathcal{O}_A$ then mainly occur due to motion of the local charge in the immediate vicinity of the boundary, leading to a boundary law.
It is important to emphasize the boundary law can also dominate in a variety of contexts beyond the low temperature limit, such as in certain excited states.

We next have to cancel out the boundary term in the second term in Eq.~(\ref{eq:comp}), which we call $\Theta_A$. 
To do so we consider a subtraction scheme based on a four-corner geometry, as illustrated in Fig.\ref{fig:corner} (right). 
Because the subregions $A,B,C$ and $D$ have an infinite boundary, the quantities $\Theta_A, \Theta_B, \dots$ are also infinite. But these boundary contributions cancel out in the following linear combination, leaving only the subleading angle-dependent correction: 
$b(\theta) = \tfrac12(\Theta_{AB}+ \Theta_{AD} - \Theta_A - \Theta_C) =  - \int_B d\bm r \int_{D} d\bm r' f(|\bm r -\bm r'|)  $. 
This integral is evaluated in Appendix \ref{AppendixA} of the Supplementary Material, where an alternative derivation, not relying on a substraction procedure, is also presented. Both methods lead to the universal corner fluctuation function \eqref{eq:master}.  
As long as the correlation function $f(r)$ decays fast enough at long distances, the radial integral in Eq.~(\ref{eq:master}) is convergent. This is guaranteed for exemple, but not exclusively, for gapped states. This integral can generally  be measured experimentally as it is directly related to  the long-wavelength limit of the static structure factor $S({\bf k})=  \int  e^{i {\bf k} \cdot {\bf r}}\braket{\rho({\bf r}) \rho(0)}_c  d{\bf r}$ associated to the observable $\mathcal{O}$:
\begin{align}
S({\bf k}\to 0) = S(0) - \pi k^2 \,\int_0^{\infty} \frac{r^3}{2} f(r) \,dr
\end{align}
Note also that $S(0) = \alpha$ gives the coefficient of the volume term.
This is natural, as bipartite fluctuations over large regions can probe the long-wavelength limit of the static structure factor \cite{Swingle_Senthil_2013}. 

We now test the super-universal shape dependence in a variety of systems, starting with quantum Hall states.

\section{Fractional quantum Hall liquids}\label{sec:fqh} 

Two-dimensional classical liquids and gapped quantum phases provide a broad and natural class of systems for which our results directly apply. In addition
  to being homogeneous and isotropic, their correlation function $f(r)$ typically decays exponentially.  An interesting example is provided by fractional quantum Hall states.  
These states are topological phases of electrons moving in two dimensions at low temperature under the influence of a strong transverse magnetic field.
They host anyon quasiparticles that are neither fermions nor bosons, and support gapless chiral edge modes.
We will study their charge fluctuations.
It is known \cite{StillingerLovett1,StillingerLovett2,Girvinetal1985,Girvinetal1986}
that for incompressible phases, the static structure factor takes the
  following form at small wavevectors:
$S({\bf k}\!\to\! 0) = l_B^2  k^2 \langle \rho \rangle/2$, where $l_B$ is the magnetic length and $\langle \rho \rangle = \nu/2\pi l_B^2$
is the electron density. The filling fraction $\nu$ gives the number of electrons per quantum
of magnetic flux. Using this result (also called a sum rule) allows us to write the full corner term:
   \begin{equation}\label{eq:sr}
b_{\nu}(\theta)=\frac{\nu}{4 \pi^2} \left( 1+(\pi-\theta)\cot \theta \right) = \frac{\sigma_{xy}}{2 \pi} \left( 1+(\pi-\theta)\cot \theta \right)\,.
\end{equation}
In the last equality, we have related
the filling fraction to the Hall conductivity in natural units, $e=\hbar=1$. 
 For the integer quantum Hall effect at $\nu \!=\! 1$, this was previously derived \cite{estienne2019entanglement}. But in fact Eq.~(\ref{eq:sr}) is valid for general incompressible interacting groundstates, including abelian and non-abelian topological states. 
 
Let us illustrate this general result with the example of the 
Laughlin state, written in first quantization as
\begin{equation}\label{eq:Laughlin}
 \Psi(z_1,\ldots,z_N)=\prod_{1\leq i<j\leq N}(z_i-z_j)^{1/\nu} e^{-\frac{1}{4}\sum_{i=1}^N |z_i|^2}\,,
\end{equation} 
for integer values of $1/\nu$.
The coordinate $z_j=x_j+i\, y_j$  of the $j$th electron is expressed as a complex number, and lengths are measured in terms of the magnetic length.
This state is a seminal example of a fractional quantum Hall fluid.
For $1/\nu \geq 2$ it has intrinsic topological order giving rise to abelian anyon quasiparticles, 
while $\nu=1$ describes the non-interacting integer quantum Hall effect. For large $N$, the particles lie in a droplet of radius $\sqrt{2N/\nu}$. In the bulk, the particle density is uniform, with exponentially decaying correlations. The conducting edge excitations are described by a chiral conformal field theory. Since quantum Hall states are gapped, fluctuation of particle number in region $A$ are expected to obey an area law. For the integer quantum Hall effect this is rigorously established~\cite{Charles_2019}, while for the fractional case this area law has been confirmed numerically~\cite{Petrescu_2014}. Furthermore in the non-interacting case the connected two-point function is known exactly so the integral (\ref{eq:master}) can be readily computed, confirming the result \eqref{eq:sr} (see Appendix \ref{Appendix_IQHE}).  However, such an elementary derivation is not viable in the interacting case, since the two-point function $f(r)$ is not known, and besides the ones we are using, only a few sum rules are known, e.g.  \cite{Sumrules_review,Kalinay2000,Zabrodin_2006,CanLaskinWiegmann,DwivediKlevtsov,Cardoso_2020}.

\begin{figure}[t]
\includegraphics[width=8.75cm]{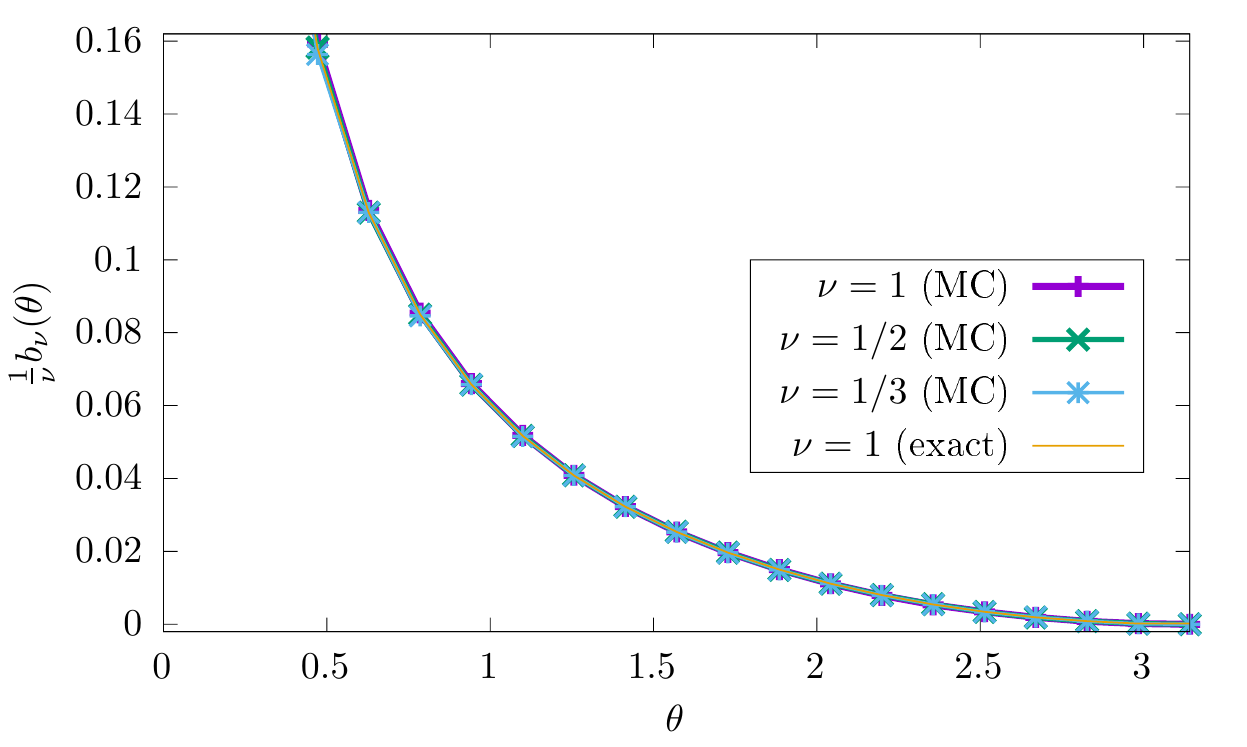}\hfill
\includegraphics[width=8.75cm]{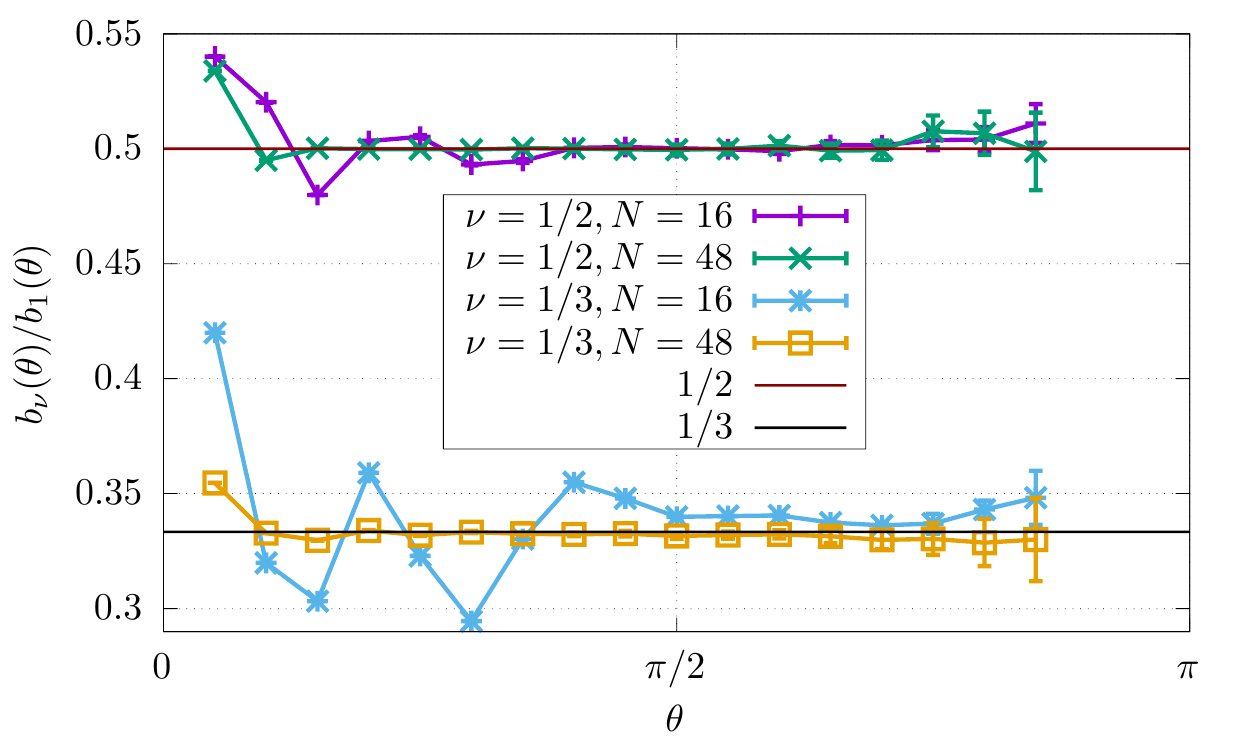}  
\caption{{\bf Charge fluctuations in fractional quantum Hall states.}
  Monte Carlo extraction of the corner term $b_\nu$ for several filling fractions. 
  {\bf Left}: $\nu^{-1}b_\nu(\theta)$ for $N=48$ particles and filling fractions $\nu=1,1/2,1/3$. The collapse onto a single curve is nearly perfect. 
  {\bf Right}: Illustration of the finite-size effects for $\nu=1/2,1/3$, by plotting $b_\nu(\theta)/b_1(\theta)$ for an increasing number of particles.
  The curves become constant for large $N$. The slight discrepancy for small $\theta$ is a finite size effect: for such small angles, there are very few particles in $A$, unless $N$ is extremely large. Note also the increase of the error bars when $\theta$ becomes close to $\pi$. This is an artifact of the fact that $b_\nu$ vanishes in that limit: while absolute error bars are still very small, relative errors blow up. } 
 \label{fig:fqhnumerics}
\end{figure}

We check the corner function \eqref{eq:master} using Monte Carlo simulations in order to sample the many-body wavefunction (\ref{eq:Laughlin}). We work with filling fractions $\nu=1/3$ and $1/2$, which correspond to topologically ordered groundstates for fermions, and bosons, respectively.
We compute the particle variance in a given subregion, with the only complication being that simulation time can become large to have a sufficient precision on the variance. Data shown in this section are typically averaged over several billion samples.
Another complication comes from the edge of the droplet, which hosts gapless chiral modes. Fortunately, the contribution from these modes is known exactly and has been shown \cite{estienne2019entanglement} to decouple from the corner contribution, so we can easily substract it (Supplementary Material). For sufficiently large particle number, the corner contribution $b_\nu(\theta)$ is given by (\ref{eq:master}), as shown in Fig.~\ref{fig:fqhnumerics}, confirming our arguments
to high precision. 

{\bf \emph{Excited states.}} We close this section by mentionning that we have verified that
super-universal shape dependence also holds for an infinite family of uniform excited states.
Let us entirely fill only the $n$th Landau level, with $n>0$, and leave all other levels empty. This is a highly excited state at unit filling (see Supplementary Material).
We find that thoses states obey the angle dependence Eq.~(\ref{eq:master}), but with $\nu$ in the prefactor replaced by 
$2n+1$. We thus see that for excited states, the prefactor is no longer simply given by the filling. It is expected that the charge fluctuations increase with the energy of the excited state. It would be interesting to understand the prefactor for other uniform excited states.

\section{Scale invariant quantum critical theories}\label{sec:cft}
After having studied gapped topological phases, we now turn to a large family of gapless systems: quantum critical phases and phase transitions.
We shall focus on systems with emergent Lorentz and scale invariance, which in the majority of cases combine to a even larger conformal symmetry. The gapless Dirac cones
of graphene, or the quantum critical transition between an insulator and superfluid at integer filling constitute key examples~\cite{subir}.  
Furthermore, the symmetries of the overarching conformal field theory impose
 the large distance behavior for the correlation function of a conserved global charge to be
$f(r) = - C_J/r^4$ \cite{Osborn1994}. $C_J$ is a positive constant that gives the universal groundstate longitudinal conductivity in natural units, $\sigma=\pi^2 C_J/2$,
of the associated conserved current. For such systems, the variance of a conserved charge obeys a strict area law (Supplementary Material). 
To find the subleading correction, we substitute this $f$ into Eq.~(\ref{eq:master}), and get
\begin{align} \label{eq:cft} 
  b(\theta)= \frac{\sigma}{\pi^2} \big( 1+(\pi -\theta)\cot\theta \big) \, \log(|\partial A|/\delta)\,.
\end{align}
The result grows logarithmically with the perimeter of $A$; we have introduced a short-distance cutoff, $\delta$.
This scaling with the perimeter is in contrast to the constant $b(\theta)$ for gapped systems. 
We note that the prefactor of the logarithm is entirely universal since it is not polluted by microscopic details (here represented by the cutoff $\delta$). 
We stress that the expression (\ref{eq:cft}) holds for any conformal field theory, irrespective of how strongly correlated it is.
Interestingly, whereas the universal Hall conductivity appeared in the corner fluctuations of quantum Hall groundstates Eq.~(\ref{eq:sr}), the above equation features the universal longitudinal conductivity,
  illustrating that universality can arise from different origins.
  In the specific case of non-interacting Dirac fermions, Eq.~(\ref{eq:cft}) was previously obtained \cite{Herviouetal2019}.

  Let us consider a different observable that is present in all CFTs, namely the energy density. The conformal symmetry constrains the two-point function to be
  $f(r) = 2 C_T/(3r^6)$ \cite{Osborn1994}, where $C_T$ is a positive coefficient that depends on the theory. As the $f$ function decays sufficiently rapidly at large distances,
  using Eq.~(\ref{eq:master}) we
  obtain a corner term with the super-universal angle dependence, with a prefactor that is constant with respect to the size of region $A$, in contrast to what was found above for
  a global charge (\ref{eq:cft}). Another difference with Eq.~(\ref{eq:cft}) is that the prefactor is no longer universal as it depends on microscopic information (the short-distance cutoff).
  In fact, we can consider infinitely many other observables, in which case the  correlation function $f$ scales as $1/r^{2\Delta}$, where $\Delta$ is the scaling dimension of the observable.
  As long as $\Delta\geq 2$, we obtain the super universal fluctuation function (\ref{eq:cft}), with a prefactor that depends on the microscopic details unless $\Delta=2$.
  This later case shows the importance of global symmetries in the study of bipartite fluctuations. In conformal quantum critical theories, there exists a small number of observables
  with $\Delta <2$, and the corresponding correlation functions decay more slowly at large distances. Interestingly, a slow decay also occurs for charge fluctuations in metals.
As we discuss next, this leads to a qualitatively distinct geometrical dependence for the fluctuations.

\section{Metals}
\label{sec:metals}
As a last example, we study fluctuations in metals. These have more mobile excitations at low energies compared to the quantum critical theories described above. For example, a two-dimensional
metal with a circular Fermi surface has an entire Fermi line worth of gapless points in momentum space, whereas Dirac semimetals only have a finite number of discrete gap-closing points. As such, it is not surprising that metals have stronger charge fluctuations 
than scale invariant critical systems. For regular metals, called Fermi-liquids, the dominant contribution to the charge fluctuations has a logarithmic enhancement compared to the boundary law, $|\partial A| \ln|\partial A|$,  with a prefactor which is known analytically \cite{Gioev_2006,Helling_2010}. This enhancement is related to the fact that $f(r)$ decays slower than for
charge fluctuations of CFTs
at large separations, namely as $1/r^3$. This has further important consequences as we now discuss.

For subsystem $A$, it is convenient to take a circular sector with radius $L$ and opening angle $\theta$. As stated before, the dominant term is a logarithmically enhanced boundary law. 
 We identified the first subleading correction, which is proportional to $L$. It is given by
\begin{align}
  b(\theta) = L\, \cffl(\theta)
\end{align}  
where an explicit formula for $\cffl$ is given in the Supplementary Material. It is different from the super-universal corner function discussed above. We stress that this term depends on the full geometry of $A$, as well as the shape of the Fermi surface. As such it becomes a full geometric term, rather than a simple corner term. Such behavior is  related to the long-range decay of the
correlation function, which blurs the notion of locality necessary to define a corner contribution. In particular, there is no reason for it to vanish at $\theta=\pi$, and be symmetric under $\theta\to 2\pi-\theta$, as before. We show in the supplementary material that this is true only up to an additive contribution, which is affine in $\theta$.
The function $\cffl$ does nevertheless contain interesting information. For example at small angles, $\cffl$ diverges logarithmically instead of the previous $1/\theta$ scaling, which further illustrates the difference with the super-univesal corner function studied in this paper. 

It would be interesting to investigate how much of this picture changes for non-Fermi-liquids, such as the fermionic half-filled Landau level \cite{HalperinLeeRead_1993}, which
can have a different decay of charge correlations.

\section{Discussion}
\label{sec:disc}
We have seen how the shape of fluctuations of an observable $\rho(\bm r)$ in a subregion with corners becomes super universal, i.e.\ it takes the same form for a very large class
of unrelated systems. In fact, the systems could be classical or quantum. We have theoretically tested our result using quantum Hall states, both fractional and integer,
and scale invariant quantum critical theories. It would be interesting to further test this super-universality in the laboratory. On the classical front, one could
study number fluctuations of colloidal particles at a two-dimensional interface (such as air/water). It should be possible to use microscopy to determine the shape dependence of the particle variance for subregions with varying corner angles. On the quantum front, a natural testbed would be ultra cold atomic gases loaded in an optical lattice.
Using various shapes of subregions one would be able to probe the atom number variance in phases like the Mott insulator or at the superfluid-to-insulator (conformal)
quantum critical point~\cite{Zhang2012,Endres2012}.     

Our analysis was mainly in two dimensions, but such super-universality is bound to occur in higher dimensions as well.
We give one concrete example in three dimensions: take subregion $A$ to be a solid cone of opening angle $\theta$, i.e.\ a 2d corner rotated about its axis of symmetry.
For simplicity, let us consider the variance of a conserved charge in scale invariant quantum critical theories described by a CFT, such as a three-dimensional Dirac semimetal.
In these CFTs, the symmetry enforces the connected correlation function to scale as $1/r^6$. Mapping the fluctuation calculation to the one for the entanglement entropy~\cite{Bueno_2019} of a special model (Supplementary Material), we find that all
such quantum critical theories will receive a correction that scales as $\tfrac{\cos^2(\theta/2)}{\sin(\theta/2)} (\log|\partial A|)^2$, with the prefactor being  
given by the universal groundstate conductivity of the system. This result is thus very similar to what we have obtained in two dimensions, Eq.~(\ref{eq:cft}). This
universal \emph{cone fluctuation} function agrees with the specific example of Dirac fermions in three dimensions~\cite{Herviouetal2019}, and holds for arbitrary CFTs. We conjecture that it will arise in the fluctuations of many other systems.
Interestingly, the cone function is the same one (up to a prefactor) that characterizes  the entanglement entropy of conical subregions in general CFTs~\cite{Klebanov2012}.  
It would be of interest to further investigate the universality of this result, and to also examine other geometries, such as trihedral corners
 appearing in polyhedra like cubes or tetrahedra.

Finally, a deeper connection between bipartite fluctuations and quantum entanglement is emerging from various directions~\cite{Gioev_2006,Songetal2012,Swingle_Senthil_2013,Herviouetal2019}, including from our super-universal results.
Indeed, the corner fluctuation function behaves almost identically as the entanglement entropy of various systems including
scale invariant quantum critical points, non-interacting Dirac fermions, quantum Hall gapped
groundstates, and supersymmetric gauge theories dual to certain string theories~\cite{Casini2008,Kallin2013,BuenoPRL,sirois2020}.     
It is an open question to understand why these distinct quantities, computed in very different systems,  
seem constrained to obey nearly the same shape dependence.

\acknowledgements 
This project was funded by a grant from Fondation Courtois, a Discovery Grant from NSERC,
a Canada Research Chair, and a ``Etablissement de nouveaux chercheurs et de nouvelles chercheuses universitaires'' grant from FRQNT.  B.E. was supported by Grant No. ANR-17-CE30-0013-01. We are grateful to Semyon Klevtsov for discussions about sum rules in fractional quantum Hall states,
and to Y.-C.~Wang, M.~Cheng, and Z.Y.~Meng for sharing their draft with us.\\ \\
\emph{Note}: In the final stage of writing, two papers that have a small overlap, specifically with our results for two-dimensional conformal field theories, appeared~\cite{wang2021,WuJianXu2021}.  
 
\clearpage
\appendix

\noindent{\LARGE \bf Supplementary Material} \\

\section{Volume and area laws in any dimension}\label{app:volarea}
In this appendix, we compute explicitly the volume and area law coefficients in any dimension $d$. While such scaling is well known for the fluctuations, we report here some general formulas in terms of the connected correlation function $f$. Our starting point is formula (\ref{eq:comp}) in the main text 
\begin{equation}
(\Delta \mathcal{O}_A)^2=  \alpha |A|  +   \Theta_A
\end{equation}
where 
\begin{equation}
 \alpha=\int d^d \mathbf{r} f(\mathbf{r}) =  \frac{2 \pi^{\frac{d}{2}}}{d\,\Gamma \left( \frac{d}{2} \right)}  \int_0^{\infty} dr\, r^{d-1} f(r) 
\end{equation}
is the coefficient of the volume term, the remaining term 
\begin{equation}
 \Theta_A= -   \int_A d^{d} \mathbf{r}_1 \int_{A^c} d^{d} \mathbf{r}_2 \, f(| \bm r_1-\bm r_2|)
\end{equation}
scales with the size of the boundary, $|\partial A|$, for large $A$:
\begin{equation}
 \Theta_A=\beta |\partial A|+\cdots
\end{equation}
where the ellipsis denote subleading terms. 
Intuitively this scaling comes from the fact that the previous integral is dominated by the region close to the interface between $A$ and $A^c$, provided $f$ decays reasonably fast. To get an explicit formula for $\beta$, we first transform the double integral over $A$ and $A^c$ into a double boundary integral over $\partial A$ using the following relation
\begin{equation}
 \Theta_A=  -   \int_{\partial A}d\sigma_1 \int_{\partial A}d\sigma_2 \, (\b n_1\cdot\b n_2)  \, F(|\b r_1-\b r_2|) 
\end{equation}
where $\bm n_1, \bm n_2$ are unit vectors normal to the boundary of $A$, and $F(r)$ is such that its laplacian satisfies $\Delta F = f$, that is $\partial_r(r^{d-1} \partial_r F(r)) = r^{d-1} f(r)$. We now pick a specific geometry where calculations are simple. In $\mathbb{R}^d$ we take for $A$ the half-space $x_d \geq 0$, with boundary $\partial A$ being the hyperplane $x_d =0$.  
\begin{equation}
 \Theta_A=   -  \int_{\mathbb{R}^{d-1}}d^{d-1} \mathbf{r}_1 \int_{\mathbb{R}^{d-1}}d^{d-1} \mathbf{r}_2   \, F(|\b r_1-\b r_2|) =|\partial A| \,  \int_{\mathbb{R}^{d-1}}d^{d-1} \mathbf{r} \, F(|\b r|)
\end{equation}
where we changed variable to the center of mass $\b r = \b r_1-\b r_2$. The fact that $|\partial A|$ is infinite is not really an issue, as one can repeat the same argument in finite volume (\emph{e.g.} working in a box with periodic boundary conditions). 
Thus
\begin{equation}
\beta=   -  \int_{\mathbb{R}^{d-1}}d^{d-1} \mathbf{r} \, F(|\b r|) =  - \frac{2 \pi^{\frac{d-1}{2}}}{\Gamma \left( \frac{d-1}{2} \right)}  \int_{0}^{\infty} r^{d-2}F(r) 
\end{equation} 
Integrating by parts twice yields
\begin{equation}
\beta= - \frac{ \pi^{\frac{d-1}{2}}}{\Gamma \left( \frac{d+1}{2} \right)}  \int_{0}^{\infty}dr\, r^{d}f(r)
\end{equation}
It is important to stress that while the computation of the area law coefficient $\beta$ has been done in this simple geometry, the  result holds irrespective of the precise shape of the boundary, unless $f$ decays too slowly.
Upon rescaling the region $A \to LA$, the fluctuations behaves for large $L$ as
 \begin{equation}
 (\Delta \mathcal{O}_{L A})^2 \sim \alpha L^d |A| +   \beta L^{d-1} |\partial A|  + \cdots
 \end{equation}
In this asymptotic regime, the boundary can be locally approximated by its tangent hyperplane, for which our computation applies.

As a side note, in two dimensions it is rather suggestive that the coefficients of the volume, boundary, and corner terms are respectively  proportional to 
\begin{equation}
\int_0^{\infty}dr\, r f(r), \qquad \int_0^{\infty}dr\, r^2 f(r) \quad \textrm{and} \quad \int_0^{\infty}dr\, r^3 f(r) \,. 
\end{equation}
Whether this remarkable sequence extends into higher dimensions is an interesting question.

\section{Two derivations of the super-universal behavior} \label{AppendixA}
In this appendix, we provide further information regarding the derivation of our main result, Eq.~\eqref{eq:master} of the main text.  
Appendix.~\ref{sec:someintegral} deals with the computation of a remaining four-dimensional integral, which gives the angular dependence of the corner function. We also present, in appendix~\ref{sec:alternativederivation}, an independent alternative derivation of the super-universal corner function.
\subsection{The remaining integral}
\label{sec:someintegral}
In this appendix we evaluate the integral 
 \begin{align}
b(\theta) =  - \int_B d{\bf r_1} \int_{D} d{\bf r_2} f(|{\bf r_1} -{\bf r_2}|)
\end{align}
This can  be done as follows. We first rewrite 
 \begin{align}
\int_B d{\bf r_1} \int_{D} d{\bf r_2} f(|{\bf r_1} -{\bf r_2}|) = \int_0^{\infty} dr f(r) \rho(r,\theta), \qquad \textrm{where} \qquad \rho(r,\theta) = \int_B d{\bf r_1} \int_{D} d{\bf r_2} \delta(|{\bf r_1} -{\bf r_2}| -r) 
\end{align}
The point is now that the regions $B$ and $D$ being cones, they are invariant under dilatations. Rescaling $r_i \to r r_i$ thus yields
 \begin{align}
\rho(r,\theta) =  \int_B d{\bf r_1}  \int_{D} d{\bf r_2}   \delta(|{\bf r_1} -{\bf r_2} | -r) =  r^3 \rho(1,\theta) 
  \end{align}
and we obtain the factorization of the angular and radial variables
\begin{align}
b(\theta) = - \rho(1,\theta) \,\int_0^{\infty} \frac{r^3}{2} f(r) \,dr \,.
\end{align}
Strikingly the angular function $\rho(1,\theta)$ does not depend on the connected density-density two-point function. The angular dependence can be
computed~\cite{estienne2019entanglement}, yielding for $\theta \in [0 , 2 \pi]$
\begin{align}
b(\theta) =  -  \left( 1+(\pi -\theta)\cot\theta \right) \,\int_0^{\infty} \frac{r^3}{2} f(r) \,dr\,.
\end{align}

\subsection{An alternative derivation}
\label{sec:alternativederivation}
For completeness, we present here an alternative derivation of our main result. Our main goal is to isolate the corner function from the dominant volume and area law term. This can be done in a different way, simply noticing that those terms are affine in $\theta$, so can be eliminated by differentiating twice with respect to $\theta$. Denoting by $D_2(\theta)$ this second derivative, we have
\begin{align}
D_2(\theta)&=\frac{d^2}{d\theta^2} \int_A d\mathbf{r}_1 \int_A d\mathbf{r}_2 f(|\mathbf{r}_1 -\mathbf{r}_2 |)\\
&=\frac{d^2}{d\theta^2} \int_0^R r_1 dr_1 \int_{0}^\theta d\theta_1 \int_0^{R} r_2 dr_2 \int_{0}^{\theta} d\theta_2 f(\sqrt{r_1^2+r_2^2-2r_1 r_2 \cos(\theta_1-\theta_2)})
\end{align}
where we integrate on an angular sector of a finite disk with radius $R$ for now. Using the identity $\frac{d^2}{d\theta^2}\int_0^\theta d\theta_1 \int_0^\theta d\theta_2 g(\theta_1-\theta_2)=g(\theta)+g(-\theta)$ and sending $R$ to infinity yields
\begin{equation}
D_2(\theta)=2\int_{0}^\infty r_1 dr_1 \int_0^\infty r_2 dr_2 f(\sqrt{r_1^2+r_2^2-2r_1 r_2 \cos \theta})
\end{equation}
which is finite. This can be evaluated by seeing $r_1$ and $r_2$ as cartesian coordinates, and switching to polar variables
\begin{align}
D_2(\theta)= \int_0^{\pi/2}d\omega \sin(2\omega) \int_{0}^{\infty} \rho^3 d\rho f(\rho\sqrt{1-\sin 2\omega \cos \theta}).
\end{align}
Finally rescaling $\rho$, we obtain
\begin{align}
D_2(\theta)&=\int_0^{\pi/2} \frac{d\omega \sin 2\omega}{(1-\sin 2\omega \cos \theta)^2}\int_{0}^{\infty} d\rho \rho^3 f(\rho)\\
&=\frac{1+(\pi-\theta) \cot \theta}{\sin^2 \theta}\int_{0}^{\infty} d\rho \rho^3 f(\rho).
\end{align}
Integrating twice, the integration constants may be set by requiring $b(\pi)=0$ and $b(2\pi-\theta)=b(\theta)$, and we recover (\ref{eq:master}).

\section{Analytic computation of the corner term for the integer quantum Hall effect} \label{Appendix_IQHE} 

In this appendix we compute the connected two-point function $f(r)$ for the $\nu = n$ integer quantum Hall effect. We obtain
      \begin{align}
      \label{eq_two_point_functio_IQH}
f(r)  =  \frac{n}{2\pi l_B^2} \delta(\bm r) -  \frac{1}{4\pi^2 l_B^4}e^{-\frac{r^2}{2 l_B^2}}\left( L_{n-1}^{(1)}\left( \frac{r^2}{2 l_B^2} \right) \right)^2\,,
   \end{align}  
   where $l_B$ is the magnetic length and $ L_{n-1}^{(1)}$ is the associated Laguerre polynomial. One can readily check that the volume term vanishes, as expected from particle number conservation. The sum rule  \eqref{eq:master} can then be computed exactly, yielding
   \begin{equation}
\int_0^{\infty} \frac{r^3}{2} f(r) dr = - \frac{n}{4\pi^2}\,.
\end{equation}

In order to derive the above relation, we first note that the corner contribution does not depend on the magnetic length $l_B$ by virtue of being dimensionless. Thus without loss of generality we set $l_B=1$. The p${}^{th}$ Landau level is spanned by the states $\ket{p,m}$ with wavefunctions (in symmetric gauge)
 \begin{align}
\Psi_{p,m}(z)  =  \frac{\sqrt{p!}}{ \sqrt{2^{m-p} m! }}   \frac{1}{\sqrt{2\pi}}  z^{m-p} L_{p}^{(m-p)}\left( \frac{z \bar{z}}{2} \right) e^{- \frac{z\bar{z}}{4}}
\end{align} 
   where $z = x+ i y$ and the integer $m$ ranges over all non negative integers. The integer quantum Hall effect  at filling fraction $\nu =n$ is obtained by occupying all Landau levels from $p=0$ to $p=n-1$. For such a non-interacting fermionic system, the connected density-density two-point function can be computed via Wick's theorem 
    \begin{align}
 \braket{\rho({\bf r_1})\rho({\bf r_2})}_c= \braket{\rho({\bf r_2})} \delta({\bf r_1}-{\bf r_2}) - \left| K \left({\bf r_1},{\bf r_2} \right) \right|^2 = K \left({\bf r_1},{\bf r_1} \right) \delta({\bf r_1}-{\bf r_2}) - \left| K \left({\bf r_1},{\bf r_2}\right) \right|^2
\end{align} 
where $K\left({\bf r_1},{\bf r_2} \right) = \braket{\Psi^{\dag}({\bf r_1}) \Psi({\bf r_2})}$ is the kernel of the projector onto the occupied states. At filling $\nu =n$, this is 
     \begin{align}
K\left({\bf r_1},{\bf r_2} \right)  = \sum_{p=0}^{n-1} \sum_{m=0}^{\infty} \Psi_{p,m}(z_1) \overline{\Psi_{p,m}(z_2) }   = \frac{1}{2\pi}e^{\frac{z_1 \bar{z}_2}{2}} e^{-\frac{z_1\bar{z}_1 + z_2 \bar{z}_2}{4}}L_{n-1}^{(1)}\left( \frac{|z_1-z_2|^2}{2} \right)
   \end{align} 
 where $z_j = x_j + i y_j$,   yielding \eqref{eq_two_point_functio_IQH}. \\
 
If we entirely fill only the $n$th Landau level for some $n \geq 0$, and leave all other levels empty, the kernel is modified to 
     \begin{align}
K\left({\bf r_1},{\bf r_2} \right)  = \sum_{m=0}^{\infty} \Psi_{n,m}(z_1) \overline{\Psi_{n,m}(z_2) }   =      \frac{1}{2\pi}e^{\frac{z_1 \bar{z}_2}{2}} e^{-\frac{z_1\bar{z}_1 + z_2 \bar{z}_2}{4}}L_{n}^{(0)}\left( \frac{|z_1-z_2|^2}{2} \right)
   \end{align} 
and we find    
 \begin{equation}
\int_0^{\infty} \frac{r^3}{2} f(r) dr = - \frac{2n+1}{4\pi^2}\,.
\end{equation}

\section{Numerical extraction of the corner term in Fractional Quantum Hall wavefunctions} \label{AppendixD} 
In the appendix, we provide more details on the numerical extraction of the corner term in the fractional quantum Hall effect. The main complication stems from the fact that the pair correlation function is not quite translational invariant for large but still finite $N$. It is in the bulk of the droplet, but there are non trivial power-law correlations at the edge. This edge behavior is well known to be described by a chiral CFT \cite{Read_2009}, and results in an extra contribution to the charge fluctuations \cite{estienne2019entanglement}. Fortunately this contribution decouples from the corner term, since correlations decay exponentially fast in the bulk.

In our geometry with opening angle $\theta$, the charge fluctuations are expected to scale as
\begin{equation}
(\Delta N_A^{\theta})^2=\alpha' \sqrt{N}+\frac{\nu}{2\pi^2}\log \left(\sqrt{N}\sin\frac{\theta}{2}\right)-b_{\nu}(\theta)+\textrm{cst}+\ldots
\end{equation}
for the Laughlin state. 
Neither the area law prefactor $\alpha'$ nor the last constant depend on $\theta$. The logarithmic term is typical for a one-dimensional CFT. 
We note that the interpretation of the factor $\nu$ is slightly different for this term, since it is the Luttinger parameter of the underlying free boson CFT. It is straightforward to extract the corner term using the above result. Provided $N$ is large enough, we have
\begin{equation}
b_{\nu}(\theta)=(\Delta N_A^{\theta})^2-(\Delta N_A^{\pi})^2-\frac{\nu}{2\pi^2}\log \left(\sin\frac{\theta}{2}\right),
\end{equation}
where $b_\nu(\theta)$, is given by (\ref{eq:master}) with the coefficient obtained in Eq.~(\ref{eq:sr}). $(\Delta N_A^{\theta})^2$ can be evaluated numerically using standard Markov chain Monte Carlo techniques, and from this we reconstruct the r.h.s.\ of the previous equation, which is shown in Fig.~\ref{fig:fqhnumerics}.

\section{Conformal Field Theories} \label{app:cft} 

For conserved charge correlation functions of CFTs in $d$ spatial dimensions, the universal large distance behavior of the correlation function is $f(r) \sim r^{-2d}$. The leading terms of the fluctuations $\Delta O_A^2$ for large regions $A$  are dominated by this infra-red behavior, including the corner terms. Thus we can ignore the short-distance behavior of $f(r)$ in evaluating $\Delta O_A^2$, at the cost of introducing a short-distance cut-off. We can transform the double integral over $A$ into a double boundary integral over $\partial A$ by virtue
of the following relation: 
\begin{align} \label{emi}
  \int_A d\b r_1 \int_A d\b r_2 \, \frac{1}{|\b r_1 -\b r_2|^{2d}} =
  -\frac{1}{2d(d-1)} \int_{\partial A}d\sigma_1 \int_{\partial A}d\sigma_2 \, \frac{\b n_1\cdot\b n_2}{|\b r_1-\b r_2|^{2(d-1)}} 
\end{align}
where $\bm n_1, \bm n_2$ are unit vectors normal to the boundary of $A$. 
Note the \emph{important minus sign} on the RHS.
Subtleties can arise due to the short-distance divergent nature of both sides, but these can be taken care of via a short-distance regulator and
do not affect the universal coefficients that interest us.
Equation (\ref{emi}) can be shown by starting with the r.h.s., and using Stokes theorem twice.  
The r.h.s.\ of (\ref{emi}) is seen to be exactly the form of the Extensive Mutual Information (EMI) model for the entanglement entropy~\cite{Casini2005,Casini2008,Swingle2010}. 
In $d=2$ spatial dimensions, the integral for region $A$ being a corner of angle $\theta$ has been computed in numerous references~\cite{Casini2008,Swingle2010,BuenoPRL}. The answer is:
\begin{align} \label{eq:stokes}
  \int_{\partial A}d\sigma_1 \int_{\partial A}d\sigma_2 \, \frac{\b n_1\cdot\b n_2}{|\b r_1-\b r_2|^2} = B\frac{L}{\delta} - a(\theta)\ln(L/\delta) + \cdots
\end{align}
where $B>0$ is the non-universal coefficient of the boundary law, and the EMI corner term reads
\begin{align}
    a(\theta) &= 2(1+(\pi-\theta)\cot\theta)
\end{align}
This leads to the universal charge fluctuation corner function of CFTs given in the main text. It is important to note that this serves as an independent derivation of our result for $b(\theta)$, Eq.~(\ref{eq:master}), for the case when $f(r)\sim 1/r^4$ at large distances.

In addition, our relation between bipartite fluctuations and the EMI for entanglement entropy gives a concrete realization of the latter,
and allows to use many of the results previously obtained in a new context.
We note that a relation similar, but distinct to (\ref{eq:stokes}) was previously obtained in Ref.~\onlinecite{Herviouetal2019}.

\section{Slow decay and metals} \label{AppendixE}
In this appendix, we focus on two-point functions that decay slower than the CFT one for a conserved charge density, which leads to different behavior for bipartite fluctuations. In particular, it is not possible anymore to interpret the term $b(\theta)$ as a corner contribution: as we shall see, this term will depend on the whole shape of region $A$. For concreteness, we consider a two-point function decaying as
\begin{equation}
 f(r)\sim \frac{a}{r^{4h}}
\end{equation}
for large $r$, and exponent $3/4\leq h<1$. For region $A$ we take a circular sector with radius $L$ and angle $\theta$. We have
\begin{align}
 \left(\Delta \mathcal{O}_A\right)^2=\int_{0}^{L} r_1 dr_1 \int_0^{L} r_2 dr_2 \int_0^{\theta}d\theta_1 \int_0^\theta d\theta_2 f(\sqrt{r_1^2+r_2^2-2r_1 r_2 \cos(\theta_1-\theta_2)})
\end{align}
To try and indentify an analog of the corner contribution, we define $D_2(\theta)$ as the second derivative of the variance. Using polar coordinates $r_1=\rho\cos\omega$, $r_2=\rho\sin\omega$, it may be expressed as
\begin{align}
 D_2(\theta)&=\int_{0}^{\pi/2}d\omega \sin 2\omega \int_{0}^{L/\cos\left(\frac{\pi}{4}-|\omega-\frac{\pi}{4}|\right)}
 d\rho\, \rho^3
 f(\rho\sqrt{1-\sin 2\omega \cos\theta})\\
 \label{eq:something}
 &\sim \frac{aL^{4-4h}}{2(1-h)}\int_0^{\pi/4} d\omega \frac{\sin 2\omega [\cos\omega]^{4h-4}}{\left[1-\sin 2\omega \cos\theta\right]^{2h}}
\end{align}
Hence, the (second derivative of the) ``corner term'' diverges as a power law in $L$. The angular dependence is no longer the super-universal function, because of the extra factor $(\cos \omega)^{4h-4}$, and the change in exponent in the denominator. The former can be traced back to the exterior (circular) boundary of $A$, which now enters the calculation due to the long-range correlation. We stress that the $\theta$ dependent correction we compute here should not be interpreted as a corner term. Indeed it is sensitive to the shape of $A$ as a whole. For instance  modifying the region $A$ even very far from the corner, such as changing the exterior boundary, does affect the angular dependence. 
Nevertheless we can consider the divergence of this correction term for $\theta\to 0$. It can be computed by noticing that the integral is dominated by the vicinity of $\omega=\pi/4$, in which case the denominator blows up. Expanding in $\omega$ up to distances of order $\sqrt{\theta}$ yields the estimate
\begin{equation}
 \int_0^{\pi/4} d\omega \frac{\sin 2\omega [\cos\omega]^{4h-4}}{\left[1-\sin 2\omega \cos\theta\right]^{2h}}\;\underset{\theta\to 0}{\sim}\;\frac{\sqrt{\pi}\,\Gamma(2h-1/2)}{\Gamma(2h)}\theta^{1-4h},
\end{equation}
where $\Gamma$ is the usual Gamma function. We also checked that the method combining four corners, which is explained in the main text, gives the same result.
Integrating twice the last integral yields a divergence as $\theta^{3-4h}$ for $h\in(3/4,1)$, and $\log \theta$ for $h=3/4$. This is to be compared with the $1/\theta$ divergence valid for any $h\geq 1$.

Let us finally discuss the case of metals. For an isotropic Fermi sea $|{\bf k}| < k_F$, the non-interacting fermion propagator is
      \begin{align}
 \braket{\Psi^{\dag}({\bf r_1}) \Psi({\bf r_2})} = K\left({\bf r_1},{\bf r_2} \right)  = \int_{|{\bf k}| < k_F} \frac{d {\bf k}}{(2\pi)^2} e^{i {\bf k} \cdot ({\bf r_1} -{\bf r_2} )}  = \frac{k_F}{2\pi r} J_1 (r k_F )
   \end{align} 
   where $r = |{\bf r_1}-{\bf r_2} |$, and $J_1(z)$ is the Bessel function of the first kind. 
   Therefore the connected two-point function is
      \begin{align}
f(r) = \frac{k_F^2}{4\pi} \frac{\delta(r)}{2\pi r} -  \frac{k^2_F}{4\pi^2 r^2} J^2_1 (r k_F )
   \end{align} 
   and up to the usual oscillations behaves at large distances as $f(r) \sim a r^{-3}$ for some constant $a$. In momentum
     space, this translates to the scaling $S(k)\propto k$ at small $k$, where $S(k)$ is the spatial Fourier transform of $f$.
   This scaling fits in the above discussion, with exponent
   $h=3/4$. For an interacting Fermi liquid (FL), the same scaling holds, but with a prefactor modified by the Landau parameter $F_{0s}$,
   as is discussed in Ref.~\onlinecite{Swingle_Senthil_2013} in the context of bipartite fluctuations.
 Therefore, for a FL $b(\theta)$ is of order $L$, $b(\theta)=L\,\cffl(\theta)$. Recall the leading term of the variance corresponds to a logarithmically
 enhanced area law (given here for non-interacting fermions)~\cite{Gioev_2006}
\begin{align}\label{eq:areaklich}
  (\Delta N_A)^2 \sim \frac{(2+\theta)}{2\pi^3}  k_F L \log L \,,
\end{align}
and this term is, indeed, eliminated by differentiating twice.
The integral in (\ref{eq:something}) can be simplified. Integrating twice, there is an ambiguity in fixing the integration constants, since changing the cutoffs---that is e.g. changing $\log L\to \log(L/\epsilon)$ in (\ref{eq:areaklich})---would result in an extra contribution of order $L$, which is affine in $\theta$. Choosing the cutoffs by asking that $\cffl(\theta)=\alpha_{\rm FL}\cdot (\theta-\pi)^2$ close to $\theta=\pi$, we obtain 
\begin{equation}
 \frac{\cffl(\theta)}{8\alpha_{\rm FL}}=\log 4-4C-2\log\left[\sin \frac{\theta}{2}\left(1+\sin \frac{\theta}{2}\right)\right]+(2\pi-\theta)\log \tan \frac{\theta}{4}+4\, \textrm{Im}\, \textrm{Li}_2\left(i\cot \frac{\theta}{4}\right).
\end{equation}
As stated above, this result is valid up to affine terms in $\theta$.   
Here, $C$ is Catalan's constant and $\textrm{Li}_2$ denotes the dilogarithm. To some extent this functional form can be seen as an analog of the corner term discussed in the text, although one should keep in mind that it is sensitive to the global geometry of the region $A$.

\bibliographystyle{apsrev4-1}
\bibliography{biblio} 

\end{document}